\documentstyle{elsart}

\begin{document}
\begin{frontmatter}
\title{Branching processes and Koenigs 
function}
\author[tch]{O.G. Tchikilev}
\address[tch]{Institute for High Energy Physics, Protvino, Russia}

\begin{abstract}
 An explicit solution of non-critical
time-homogeneous    branching processes  is described.
\end{abstract}
\end{frontmatter}

\section{Introduction}
 Branching processes are widely used in high energy physics~\cite{hwa1}. 
 For example, the well known Furry-Yule and 
 negative binomial distributions occur in simple
 branching processes with allowed transition $1 \rightarrow 2$. The use of
 processes with higher order transitions $1 \rightarrow n$ with $n>2$  is
 rare due to the absence of explicit solutions in terms of elementary functions.
 In this paper we describe the solution for  processes with higher
 order transitions using a
 recently found recursive procedure for the pure birth
 branching process\cite{tch1}. The solution is based on the use of the
 Koenigs function\cite{kon1} and the functional Schr\"oder equation\cite{sch1}, 
  sometimes called the Schr\"oder-Koenigs equation 
 (for a detailed description and extensive bibliography see~\cite{kuc1,kuc2}).
  In section~2 we describe the solution\cite{tch1} for the pure birth 
 branching process. In sections 3 and 4, respectively, the procedures
 for non-critical branching processes and branching processes
 with immigration are outlined. The results
 are summarized in the last section.
\section{Solution for the general pure birth branching process} 
 A branching process with continuous evolution
 parameter $t$ is determined by the  rates $\alpha_n$ for the transition 
(``splitting'') of one particle into $n$ particles with all particles 
 subsequently
 evolving independently. For a pure birth branching process
 $\alpha_0 =0$.
The probability distribution
 $p_n(t)$ for the process having one particle at $t=0$ with
$ p_n(0) = \delta _{1n} $ is a solution of the forward Kolmogorov 
 equation~\cite{kol1,kol2}
\begin{equation}\label{eq:1}
  \frac{\partial m}{\partial t} = f(x) \frac{\partial m}{\partial x}
\end{equation}
 for the probability generating function
\begin{equation}
 m(x,t) = \sum _{n=0}^{\infty} p_n (t) x^n
\label{eq:01}
\end{equation}
 with
\begin{equation}\label{eq:2}
 f(x) = \sum_{n=2}^{\infty} \alpha_n x^n - \alpha x 
\end{equation}
 and with
 $\alpha = \sum \alpha_n$.
 The Taylor expansion of  equation~(\ref{eq:1}) leads to the following
 system of equations for the probabilities $p_n$
\begin{eqnarray}
 \frac{dp_1}{dt} = - \alpha p_1 \, \qquad , \\
 \frac{dp_2}{dt} =  \alpha_2 p_1 - 2 \alpha p_2 \, \qquad , \\
 \frac{dp_3}{dt} = \alpha_3 p_1 +2 \alpha_2 p_2 - 3 \alpha p_3\, 
\label{eq:4}
\end{eqnarray}
 and for arbitrary $n$
\begin{equation}
 \frac{dp_n}{dt} = \sum_{j=1}^{n-1} j \alpha_{n-j+1} p_j - n \alpha p_n
 \qquad .
\label{eq:5}
\end{equation} 
 A simple interpretation of the equation (\ref{eq:5}) is as follows: let us
 consider the state with $n$ particles at the moment t. The change in this
 state is due to the arrival from states with multiplicity lower than $n$ and to
 the departure to states with higher multiplicity. The arrival rate
 from the state with $j$ particles is proportional to the number of particles
 $j$, to the transition rate (for one particle) to produce $n-j$ new particles,
 i.e. $\alpha_{n-j+1}$, and to the population density in the state $j$, the sum
 in the equation (\ref{eq:5}) goes over all states below $n$. The departure rate
 is proportional to the total transition rate (for one particle) $\alpha$,
 to the number of particles $n$ in this state and to the population density, 
 this
 explains second term in the equation (\ref{eq:5}). Formally this system of
 equations is valid for any initial condition.
 
 Let us recall the recursive solution of the equations (4)-(7)
 given in \cite{tch1}: the probability $p_1(t)$ is
$  p_1 = \exp ( -\alpha t )$ and
\begin{equation}
 p_n = \sum _{j=1}^{n} \pi _{jn} p_{1}^{j}
\label{eq:7}
\end{equation}
 with the following recursion for the coefficients $\pi_{jn}$
\begin{equation}
 (n-j)\pi_{jn} = \sum _{l=1}^{n-j} (n-l)b_l~ \pi_{j(n-l)} \qquad .
\label{eq:8}
\end{equation}
 Here $ b_l = \alpha_{l+1} / \alpha $ is the relative probability to produce
 $l$ new particles. The recursion starts from $\pi_{11} = 1$ and the
 coefficient $\pi_{nn}$ can be found from the initial condition 
\begin{equation}
 \pi_{nn} = - \sum_{j=1}^{n-1} \pi _{jn} \qquad .
\label{eq:9}
\end{equation} 
 For the case with N initial particles with 
 $ p^{(N)}_n (0) = \delta _{Nn} $, 
 the first N-1 equations are automatically valid since
\begin{equation}
 p_1^{(N)}=p_2^{(N)}=...=p_{N-1}^{(N)}=0 
\label{eq:11}
\end{equation}
 and the  solution in this case has the following form:
$ p^{(N)}_N = \exp ( -N\alpha t) = p_1^N $  and
\begin{equation}
 p^{(N)}_n = \sum _{j=N}^{n} \pi^{(N)}_{jn} p_1^j
\label{eq:13}
\end{equation}
 with the same recursion as (\ref{eq:8}) for the coefficients $\pi ^{(N)}_{jn}$
\begin{equation}
 (n-j)\pi ^{(N)}_{jn} = \sum _{l=1}^{n-j} (n-l) b_l~ \pi^{(N)}_{j(n-l)} \qquad .
\label{eq:14}
\end{equation}
 This recursion starts from $\pi^{(N)}_{NN} =1$ and the coefficient
 $\pi^{(N)}_{nn}$ can be found from the relation
\begin{equation}
 \pi^{(N)}_{nn} = - \sum _{j=N} ^{n-1} \pi^{(N)}_{jn} \qquad .
\label{eq:15}
\end{equation}
 One can calculate the coefficients $\pi^{(N)}_{jn}$ using the concept of the
Koenigs function~\cite{tch1,kuc1,kuc2,val1}. For the branching process
starting from one particle at $t=0$ this function is defined as the limit
\begin{equation}
 K(x) = \lim_{n\rightarrow\infty} \frac{m(x,nt)}{p_1^n} \qquad .
\label{eq:16}
\end{equation}
  $K(x)$ has the following Taylor expansion:
\begin{equation}
 K(x) = \sum_{j=1}^{\infty} \kappa_j x^j = \sum_{j=1}^{\infty} \pi_{1j} x^j
 \qquad .
\label{eq:17}
\end{equation}
 For the branching process starting from N particles the Koenigs function is
 defined analogously:
\begin{equation}
 K^{(N)}(x) = K^N(x) = \lim_{n\rightarrow\infty} \frac{m^N(x,nt)}{(p_1^N)^n} =
 \sum_{j=N}^{\infty} \kappa_j^{(N)} x^j = 
 \sum_{j=N}^{\infty} \pi_{Nj}^{(N)} x^j
 \qquad . 
\label{eq:017}
\end{equation}
The recursion (\ref{eq:14}) leads to the following recurrence for the
coefficients $\kappa^{(N)}_{j}, N=1,2,...; ~j=N+1, N+2, ...$~:
\begin{equation}
 (j-N)\kappa_j^{(N)} = \sum_{l=1}^{j-N} (j-l) b_l~ \kappa^{(N)}_{(j-l)} \qquad.
\label{eq:18}
\end{equation} 
 Let us denote
$ \kappa ^{(x)}_{x+n} = t_n (x)$ ,
 then $t_0 (x) =1$ and $t_n(x)$ is given by the following recursion
\begin{equation}
 n t_n (x) = \sum_{l=1}^{n} (x+n-l) b_l~ t_{n-l} (x) \qquad .
\label{eq:20}
\end{equation}
 It is evident from the equation (\ref{eq:20}) 
 that the $t_n (x)$ is a polynomial of order $n$  in $x$.
 The $\kappa^{(N)}_j$ in terms of $t_n(x)$ is equal to $t_{j-N} (N)$. 

 The remarkable property of the Koenigs function is
 that it satisfies the functional Schr\"oder
equation:
\begin{equation}
 K(m) = p_1 K(x)
\label{eq:21} \qquad.
\end{equation}
 It is convenient to introduce the function $Q(x)$ the inverse of the Koenigs
 function. Then the equation (\ref{eq:21}) gives the functional
 relations
 $m(x,t)=Q(p_1 K(x))$ and
 $  m^N(x,t) = Q^N (p_1 K(x)) $ .
 The coefficients of the Taylor expansion for $Q^N(x)= \sum Q^{(N)}_j x^j$ 
can be found using
the following relation (see Appendix~B in \cite{traub1} and references 
therein):
\begin{equation}
 Q^{(N)}_j = \frac{N}{j} \kappa ^{(-j)}_{-N} \qquad .
\label{eq:23}
\end{equation}
 In terms of $t_n(x)$ it gives
\begin{equation}
 Q^{(N)}_j = \frac{N}{j} t_{j-N} (-j) \qquad .
\label{eq:24}     
\end{equation}
 Finally,  comparison of  equation (\ref{eq:13}) with the Taylor 
 expansion in $x$ of the $Q^N(p_1(K(x))$
leads to the following expression
 for the coefficients $\pi^{(N)}_{jn}$:
\begin{equation}
 \pi ^{(N)}_{jn} = Q^{(N)}_j \kappa^{(j)}_n = 
\frac{N}{j} t_{j-N}(-j)t_{n-j}(j)  \qquad .
\label{eq:25}
\end{equation}
\section{Solution for non-critical branching processes}
In this case the absorption coefficient $\alpha_0$ is not equal to zero and
the function $f(x)$ (\ref{eq:2}) has an additional term $\alpha_0 (1-x)$.
 Let us denote by $\beta$ the smallest positive, different from unity 
root of the equation $f(x)=0$. Branching processes with $0 < \beta < 1$
are called supercritical branching processes and processes with $1 < \beta$
are called subcritical ones.

 The solution for the supercritical branching processes is obtained in the
following way. Let us transform $m$ and $x$ to $m'$ and $x'$ by the linear
transform
\begin{equation}
 x' = \frac {x-\beta}{1-\beta}~~ , \qquad  
 m' = \frac {m-\beta}{1-\beta} \qquad ,
\label{eq:26}
\end{equation}
 this transform moves the root $\beta$ to zero. Then the forward Kolmogorov
 equation (\ref{eq:1}) for the process with absorption
transforms to the form for the pure birth branching process.
 Therefore one can introduce the Koenigs function with
\begin{equation}
 K \biggl(\frac{m-\beta}{1-\beta}\biggr) = 
 q_1 K \biggl(\frac{x-\beta}{1-\beta}\biggr) \qquad ,
\label{eq:27}
\end{equation}
 where $q_1 = \exp {(-\alpha' t)}$~.
 This leads to the expression for the $m(x,t)$
\begin{equation}
 m(x,t) = \beta + (1-\beta) Q\biggl(q_1 
 K \biggl(\frac{x-\beta}{1-\beta}\biggr)\biggr)
\label{eq:28}
\end{equation}
and to the infinite series in $q_1$ for the probabilities $p_n$
\begin{equation}
 p_n = \beta \delta_{0n} + (1-\beta) \sum_{l=n}^{\infty}
 \frac{(-\beta)^{l-n}}{(1-\beta)^l} \frac {l!}{n!(n-l)!}
 \sum_{j=n}^{l} Q_j q_1^j \kappa^{(j)}_l \qquad .
\label{eq:29}
\end{equation}

 The solution for the subcritical branching processes ($\beta>1$) 
 can be obtained in a similar way.
  In this case the linear transform should move 1 to zero and 
 $\beta$ to 1, i.e. $x' = (x-1)/(\beta-1)$. This leads to the similar
 expressions
\begin{equation}
 m(x,t) = 1 + (\beta-1) Q\biggl(q_1 K\biggl(\frac{x-1}{\beta-1}\biggr)\biggr)
\label{eq:30}
\end{equation}
 and
\begin{equation}
 p_n = \delta_{0n} + (\beta-1) \sum_{l=n}^{\infty} \frac{(-1)^{l-n}}
 {(\beta-1)^l} \frac{l!}{n!(n-l)!}
 \sum _{j=n}^{l} Q_j q_1^j \kappa^{(j)}_l \qquad .
\label{eq:31}
\end{equation}

 The procedures described above are not suitable for critical branching 
 processes with $\beta=1$. In simplest case without higher order transitions 
 the probability generating function $m(x,t)$ is a linear fraction with 
 coefficients having linear dependence on $t$ (instead of $\exp (-\alpha t)$ 
 dependence for non-critical processes). In general case 
 with higher order transitions the interplay between
 $t$ and $\exp (-\alpha t)$ dependences leads to more complex equations.

\section{Solution for branching processes with immigration}
 For the branching processes with immigration there is an additional external
 source of particles appearing in clusters of $j$ particles with the
 differential
 rates $\beta_j$ ( $\sum \beta_j = b$ ). The generating function for the
 process starting with zero particles at $t=0$ can be written~\cite{bar1,tch2}
 as
\begin{equation}
 M(x,t) = \exp \biggl( \int_{0}^{t} g(m(x,\tau )) d\tau \biggr)
\label{eq:32}
\end{equation}
 with
\begin{equation}
 g(x) = \sum_{i=1}^{\infty} \beta_i x^i - b  \qquad ,
\label{eq:33}
\end{equation}
 where $m(x,\tau)$ is the solution for the underlying branching
 process without immigration.  For the underlying pure birth branching
 process,  equation (\ref{eq:32}) leads to the following expression
\begin{equation}
 M(x,t) = \exp {(-bt)} \exp \biggl( \sum_{n=1}^{\infty} v_n(t) x^n \biggr)
\label{eq:34}
\end{equation}
 with
\begin{equation}
 v_n(t) = \sum_{i=1}^{n} \sum_{j=i}^{n} \pi^{(i)}_{jn} \frac{1-p_1^j (t)}
 {j\alpha} \qquad  .
\label{eq:35}
\end{equation}
 Let us denote
\begin{equation}
 \exp {( \sum_{n=1}^{\infty} v_n x^n )} = 1 + \sum_{n=1}^{\infty} V_n x^n 
 \qquad ,
\label{eq:36}
\end{equation}
 then the final probability $P_n(t) =\exp {(-bt)} V_n(t)$. The coefficients
 $V_n(t)$ can be calculated using the recursive relation:
\begin{equation}
  n V_n = \sum _{j=1}^{n} j v_j V_{n-j} \qquad.
\label{eq:37}
\end{equation}
 This relation is known in combinatorics and is used, for example, in the
 study of combinants\cite{comb1,comb2,comb3,comb4}.
\section{Summary}
  In this paper we have derived explicit expressions for the  probability 
 distributions in various branching processes. Although we have not derived 
 closed expressions for the polynomials $t_n(x)$,  the given recursions 
 can serve as  a calculational
 tool both in theoretical and experimental studies. 

\end{document}